\documentclass[12pt, a4paper, twoside]{article}
\usepackage{amsmath,amsfonts,amssymb,amsthm,amscd}
\usepackage{epsfig}
\usepackage{graphicx}
\usepackage[cp1250]{inputenc}
\usepackage{hyperref} 
\usepackage{enumerate} 

\newtheorem{theorem}{Theorem}

\newtheorem{lemma}[theorem]{Lemma}
\newtheorem{corollary}[theorem]{Corollary}

\def\l{\left}
\def\r{\right}
\def\epsi{\varepsilon}

\def\rdim #1{\mathbb R^{#1}}

\def\diam{\mathrm{diam}}
\def\fix{\mathrm{fix}}
\def\prob{\mathrm{Prob}}

\def\mycirc{\mathcal S} 
\def\hb{H_{\mathrm{b}}}
\def\hr{H_{\mathrm{r}}}
\begin{document}

\title{Untangling polygons and graphs \\[20pt] 
}

\author{
{Josef Cibulka}\thanks{Supported by project MSM0021620838 of the Czech Ministry of Education.}  \\[-2mm]
{\small\it Department of Applied Mathematics}    \\[-2mm]
{\small\it Charles University}                \\[-2mm]
{\small\it Malostransk\'e n\'am.~25}          \\[-2mm]
{\small\it 118~00 Prague, Czech Republic}    \\[-2mm]
{\small\tt cibulka@kam.mff.cuni.cz}           \\[-2mm]
}


\date{}
\maketitle

\normalsize 

\begin{abstract}
Untangling is a process in which some vertices of a planar graph are moved to obtain
a straight-line plane drawing. The aim is to move as few vertices as possible.
We present an algorithm that untangles the cycle graph $C_n$ while keeping at least $\Omega(n^{2/3})$
vertices fixed. For any graph $G$, we also present an upper bound on the number of fixed 
vertices in the worst case.
The bound is a function of the number of vertices, maximum degree and diameter of $G$.
One of its consequences is the upper bound $O((n \log n)^{2/3})$ for all 3-vertex-connected
planar graphs.

\end{abstract}

\section{Introduction}
Given any planar graph whose every vertex has a prescribed position in the plane,
it is possible to draw the graph so that edges are pairwise non-crossing curves.
According to F\'ary's theorem~\cite{Fary1948}, 
every planar graph can also be drawn in the plane so that edges are 
non-crossing straight line segments. However, to obtain a straight-line plane drawing 
of a graph with given vertex positions, we may need to move some of the vertices; 
this process is called \emph{untangling}. It is natural to ask at most how many vertices 
can keep their positions during untangling. If a vertex keeps its position, it is 
called \emph{fixed}, otherwise it is called \emph{free}.

In the following, a \emph{drawing} of a graph will always mean a straight-line drawing
in $\rdim 2$, which is completely determined by the positions of the vertices. A drawing is 
\emph{plane} if no two edges cross.

Let $G$ be a graph and let $\delta$ be a mapping of vertices of $G$ to points in the plane. 
We define 
\begin{eqnarray*}
\fix(G,\delta) &=& \max_{\beta \text{ plane drawing of } G} \{|v \in V(G): \beta(v)=\delta(v)|\}, \\
\fix(G) &=& \min_{\delta \text{ mapping of }V(G)\text{ to }\rdim 2} \{\fix(G,\delta)\}.
\end{eqnarray*}

At the 5th Czech-Slovak Symposium on Combinatorics in Prague in 1998, Mamoru Watanabe asked
whether every polygon on $n$ vertices can be turned into a noncrossing polygon by moving at 
most $\epsi n$ its vertices for some constant $\epsi>0$. This is equivalent to asking whether 
$\fix(C_n)\geq (1-\epsi) n$ for the cycle graph $C_n$.
Pach and Tardos~\cite{PachTardos} answered this question in the negative by showing 
$\Omega(\sqrt{n}) \leq \fix(C_n)\leq O((n \log n)^{2/3})$.
We almost close the gap for cycle graphs by designing an algorithm 
that always keeps at least $\Omega(n^{2/3})$ vertices fixed.

The following table summarizes the best known bounds on $\min_{G\in\mathcal G, |V(G)|=n}(\fix(G))$ for 
several graph classes $\mathcal G$.

\bigskip
\begin{tabular}{c|cc|cc}
Graph class $\mathcal G$& Lower bound & & Upper bound & \\ 
\hline
Cycles & $\Omega(n^{2/3})$ & \autoref{thm:cyc_lb} & $O((n \log n)^{2/3})$ & \cite{PachTardos} \\ 
\hline
Trees & $\l\lceil \sqrt{n/2} \r\rceil$ & \cite{Goaoc+2008} & $3\sqrt{n}-3$ & \cite{Bose+2007} \\ 
\hline
Outerplanar graphs & $\sqrt{(n-1)/3}$ & \cite{SpillnerWolff} & $2\sqrt{n-1} + 1$ & \cite{SpillnerWolff} \\ 
\hline
Planar graphs & $\sqrt[4]{n/3}$ & \cite{Bose+2007} & $\l\lceil \sqrt{n-2} \r\rceil + 1$ & \cite{Goaoc+2008} \\ 
\end{tabular}
\bigskip

It is known, that $\fix(G) \geq \Omega(n^{1/4})$~\cite{Bose+2007} and  
$\fix(G) \geq \Omega(\sqrt{\Delta(G)}+\sqrt{\diam(G)})$~\cite{SpillnerWolff}, where $\Delta(G)$ is
the maximum degree and $\diam(G)$ is the diameter of the given planar graph $G$ on $n$ vertices.

In \autoref{sec_ub}, we present a general upper bound on $\fix(G)$ of a 
planar graph $G$. The bound is a function of the number of vertices, 
maximum degree and diameter of $G$. This general upper bound has three interesting special
cases:

\begin{itemize}
\item 
For any 3-vertex-connected planar graph $G$, $\fix(G) \leq O((n \log n)^{2/3})$.
\item
The upper bound 
$O(\sqrt{n} (\log n)^{3/2})$ for any planar graph such that its maximum degree and diameter
are both in $O(\log n)$. This is close to the lowest known
upper bound on $\fix(G)$ for some graph $G$, which has value $O(\sqrt{n})$, but was
established only for several special graphs~\cite{Bose+2007, Goaoc+2008, SpillnerWolff}.
\item
For any constant $\epsi$, if a graph $G$ satisfies 
$\fix(G)\geq \epsi n$, then it has a vertex of degree at least $\Omega(n\epsi^2 /\log^2 n)$.
\end{itemize}

All logarithms in this paper are base 2.

\section{Algorithm for untangling cycles}

Let $C_n$ be the graph with vertices $v_1, v_2, \dots, v_n$ and edges 
$(v_1,v_2)$, $(v_2,v_3)$, \dots, $(v_n,v_1)$.
\begin{theorem}
\label{thm:cyc_lb}
$ \fix(C_n) \geq 2^{-5/3}n^{2/3} - O(n^{1/3}) = \Omega(n^{2/3}) $
\end{theorem}

\begin{proof}
Let $m$ be the largest integer such that $m \leq n-4$ and $(m/16)^{1/3}$ is an integer. Then 
$m\geq n-O(n^{2/3})$. Define $l:=(m/16)^{1/3}$ and $s := 2 l$. For the given vertex positions,
we fix a horizontal direction so that no two vertices lie on any horizontal or vertical line. 

Vertices $v_{m+1}, v_{m+2} \dots v_n$ will be free and will make bends on the line between the 
last and the first fixed vertex. 
We will divide some of the remaining vertices to $2l$ layers, each with $s^2$ vertices. 
The first layer consists of the highest $s^2$ vertices. Using the Erd\H{o}s-Szekeres 
lemma~\cite{ErdosSzekeres}, we select among them
a sequence of exactly $s$ vertices with indices either increasing or decreasing from left to 
right. There are two types of layers --- if the selected vertices have increasing indices, 
the layer is an \emph{increasing layer}, otherwise it is a \emph{decreasing layer}. 
After we select the monotone sequence of vertices, we free all the remaining vertices of the layer.
We also free all the vertices that are below this layer and are at graph distance at most $2(l-1)$ 
from some of the selected vertices.

In general, each layer consists of the highest $s^2$ vertices that are not free 
and lie below all previous layers. From the layer we select a monotone sequence 
of length $s$ and free the remaining vertices of the layer. Then 
we free every vertex that lies below this layer and whose index differs by at most $2(l-i-1)$ 
from index of some of the selected vertices. Here $i$ is the number of previously created layers of 
the same type.

We need to count that we have enough vertices for all the layers. Each of the $2l$ layers 
consists of $s^2$ vertices and for each of the $s$ selected vertices in the $i^{\mathrm{th}}$ 
increasing or in the $i^{\mathrm{th}}$ decreasing layer, we freed at most $4(l-i)$ vertices
lying below the layer. The number of considered vertices is thus at most
\[
2ls^2 + 2s\sum_{i=1}^{l}(4(l-i)) = 8l^3 + 4sl(l-1)) \leq 16l^3 = m .
\]

Without loss of generality, we have $l$ increasing layers, each having $s$ selected vertices whose 
indices increase from left to right. These $ls$ vertices are our fixed vertices; call them 
$u_1 = v_{i_1}$, $u_2 = v_{i_2}$ \dots $u_{ls} = v_{i_{ls}}$, where $i_1 < i_2 < \dots < i_{ls}$. 
We assign new positions of the free vertices $u_j$ satisfying $i_1<j<i_{ls}$ in the order of 
increasing indices.
If $u_i$ and $u_{i+1}$ lie in the same layer,
we connect them by a straight line segment and place on it the free vertices between $u_i$
and $u_{i+1}$. Otherwise, we have at least $2d+2$ free vertices between $u_i$ and $u_{i+1}$,
where $d$ is the number of increasing layers between layers 
of $u_i$ and $u_{i+1}$. We will view the path between $u_i$ and $u_{i+1}$ as a line formed by
straight line segments and $2d+2$ bends. All the segments will be either horizontal or vertical, 
except for the first one, which goes from $u_i$ almost vertically and slightly to the right to avoid 
having a common subsegment with the last segment of the path between $u_{i-1}$ and $u_i$. 
Each vertical segment passes through 
one layer so that all segments already placed in this layer are to the left of it and all the fixed 
vertices $u_j$ in this layer and with $j>i$, are to the right. The horizontal segments are 
placed between layers to connect pairs of vertical segments. At the end, we will connect 
$u_{ls}$ and $u_1$ by a line with 4 bends and place on it all the remaining free vertices. 
See \autoref{fig_join}.

\begin{figure}[hbt]
\begin{center}
	\epsfbox{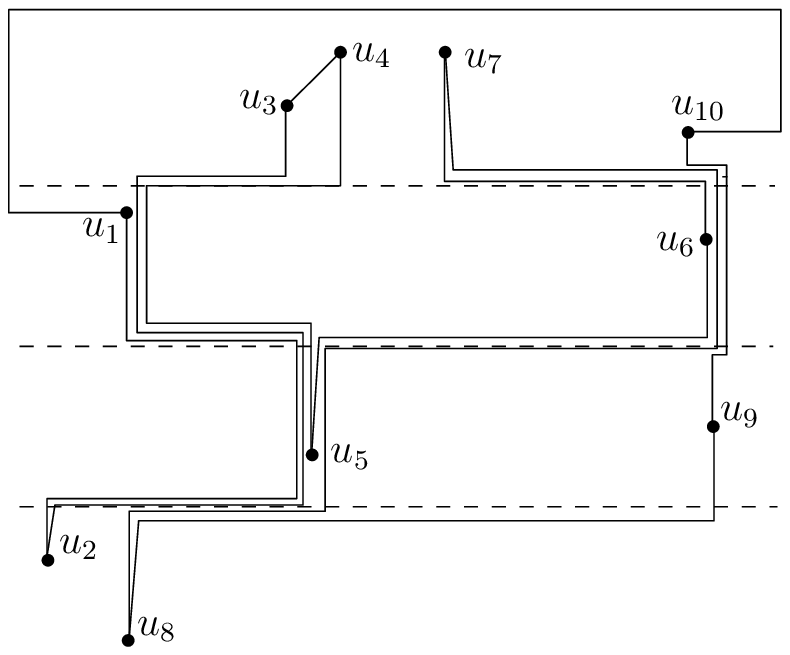}
	\caption{joining the increasing sequences}
	\label{fig_join}
\end{center}
\end{figure}

The number of fixed vertices is
\[
ls = 2l^2 = 2^{-\frac53}m^{\frac23} \geq 2^{-\frac53}n^{\frac23} - O(n^\frac13) .
\]
\end{proof}

\section{Upper bounds}
\label{sec_ub}

We will use the following lemma of Pach and Tardos~\cite{PachTardos} who used it to show 
the upper bound $\fix(C_n) = O((n\log n)^{2/3})$.

\vbox{
\begin{lemma} 
\label{lemma_pt}
(Pach and Tardos 2002 \cite{PachTardos}) \par
Let $\hb$ be the graph with vertices $u_1,u_2, \dots, u_t$ and edges 
$(u_1,u_2)$, $(u_2,u_3)$, \dots, $(u_t,u_1)$.
Let $\hr$ be a random Hamiltonian cycle on the same vertex set, that is, its edges are
$(u_{\sigma(1)},u_{\sigma(2)})$, $(u_{\sigma(2)},u_{\sigma(3)})$, \dots $(u_{\sigma(t)},u_{\sigma(1)})$,
where $\sigma$ is a random permutation of $\{1,2,\dots,t\}$.

Then, for any integers $s<t$ and $K$, the probability that the crossing number of \hbox{$H=\hb \cup \hr$}
is at most $K$ satisfies
\[
\text{Prob}[\text{cr}(H) \leq K] \leq \binom{t}{D}^2 \l(\frac{2t}{s}\r)^D \frac{s^{t-D}}{(t-D)!} ,
\]
where $D := \l\lfloor 35\sqrt{t(K+t)/s}  \r\rfloor$.
\end{lemma}
}

The edges of $\hr$ and $\hb$ are called \emph{red edges} and \emph{black edges}, respectively. 

\begin{theorem}
\label{thm_ub}

Let $T$ be a tree on $n$ vertices. Let $\Delta$ be the maximum degree of $T$ 
and let $\diam$ be the diameter of $T$. Then
\[
\fix(T) \leq 300 \sqrt{n} \log n \l(\sqrt{\Delta} + \min\l\{\sqrt[6]{n/\log^2 n},~ \sqrt{\diam} \r\}\r) .
\]
\end{theorem}

\begin{proof}
If $n \leq 90000$, the statement is trivially true, so we will assume $n>90000$.

Let $T$ be a given tree on $n$ vertices. We will select one of its vertices to be the root.
Let a \emph{DFS-cycle} be an oriented closed walk that follows some 
depth-first search (DFS) performed on $T$ starting in the root. A DFS-cycle contains each orientation
of every edge exactly once. By a \emph{drawing of a DFS cycle} we will mean a drawing (not necessarily 
straight-line) of the oriented cycle graph whose vertices and edges correspond to occurrences of vertices 
and edges on the DFS-cycle. Vertices of the drawing corresponding to a vertex $v$ of $T$ are placed near 
to the position of $v$.

\begin{figure}[hbt]
\begin{center}
	\epsfbox{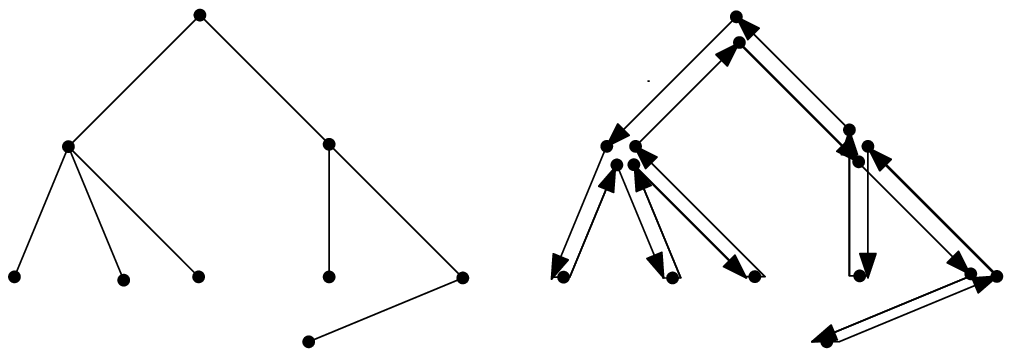}
	\caption{a tree and a drawing of one of its DFS-cycles}
	\label{fig:DFS-cyc}
\end{center}
\end{figure}

We fix one DFS-cycle and number the vertices in the order in which 
they were first visited during the depth-first search; the vertices will be $v_1, v_2, \dots, v_n$.
Then we place all the vertices of $T$ in a random order to the vertices of a convex polygon. 
This mapping of vertices to points will be $\delta$. There are $\binom{n}{t}$ 
possible selections of $t$ fixed vertices. The fixed vertices will be $u_1 = v_{i_1}$, \dots 
$u_t = v_{i_t}$, where $i_1 < i_2 < \dots < i_{t}$. For each such selection, let $\hb$ be defined 
as in Lemma~\ref{lemma_pt} and let $\hr$ be the cycle whose edges are the edges of the 
convex hull of $u_1$, $u_2$, \dots, $u_t$. 

In the rest of the proof we will find a drawing of $H$ with small crossing number
assuming that $T$ can be untangled while keeping $u_1,\dots, u_t$ fixed. We will then use 
Lemma~\ref{lemma_pt} to show that the probability that this happens for at least one selection of
$u_1,\dots, u_t$ is smaller than one. Thus there will be an ordering of the vertices of $G$
such that $G$ cannot be untangled while keeping $t$ vertices fixed.

The cycle $\hr$ is drawn as the convex hull of $u_1$, $u_2$, \dots, $u_t$. Thus there are no 
crossings of pairs of red edges.

Let $T'$ be the smallest subtree of $T$ that contains all the fixed vertices. 
We take the DFS-cycle on $T'$ that comes from the fixed DFS-cycle on $T$ by omitting 
vertices outside $T'$.
To obtain a drawing of $\hb$ we will find a drawing of the DFS-cycle on $T'$ and then each 
edge $(u_i,u_{i+1})$ of $\hb$ will be drawn as the part of the DFS-cycle between the first 
occurrences of $u_i$ and $u_{i+1}$.

We take the plane drawing of $T$ in which $u_1, \dots, u_t$ keep their positions. 
From this drawing, we get a drawing of the DFS-cycle on $T'$ by expanding every vertex of $T'$ 
as in \autoref{fig_tree_expand}:

We start by expanding the root of the DFS. We then expand each vertex $v_i$ when 
its only expanded neighbor is its parent (that is, its unique neighbor on the path to the root).
Let $p_i$ denote the parent of $v_i$; if $v_i$ is the root, then we define $p_i$ to be the second
vertex visited during DFS. 
We take a point $C$ near to $v_i$ on the line joining $p_i$ and $v_i$, but outside the segment 
between $p_i$ and $v_i$. 
Let $\mycirc$ be a circle centered at $C$ with radius chosen so that $v_i$ lies on $\mycirc$. 
Vertex $v_i$ will be expanded to several sub-vertices placed on $\mycirc$. 
First, we create two sub-vertices:
$\hat{v_i}$ which keeps the position of $v_i$ and $\bar{v_i}$ which is placed near to $\hat{v_i}$.
Sub-vertices $\hat{v_i}$ and $\bar{v_i}$ are ends of the straight line segments 
corresponding to the oriented edges $(p_i,v_i)$ and $(v_i,p_i)$ of the DFS-cycle, respectively.
Vertex $\bar{v_i}$ is placed so that the segments corresponding to $(v_i,p_i)$ and $(p_i,v_i)$ do not 
cross. For every $v_j$ neighbor of $v_i$ different from $p_i$, we create sub-vertices $\hat{v}_i^j$
and $\bar{v}_i^j$ near the intersection of $\mycirc$ and the segment between $C$ and $v_j$. 
Sub-vertices $\hat{v}_i^j$ and $\bar{v}_i^j$ are ends of the segments corresponding to 
the oriented edges $(v_j,v_i)$ and $(v_i,v_j)$ of the DFS-cycle, respectively.
Finally, we connect pairs of the newly created sub-vertices by segments inside $\mycirc$ ---
if $(v_j,v_i)$ is followed by $(v_i,v_k)$ in the DFS-cycle, then we connect $\hat{v}_i^j$ with 
$\bar{v}_i^k$ (if $v_j=p_i$ then we define $\hat{v}_i^j := \hat{v_i}$ and similarly if $v_k=p_i$
then $\bar{v}_i^k := \bar{v_i}$). 
If $v_i$ has degree two, then we place the sub-vertices $\hat{v}_i^j$ and $\bar{v}_i^j$ so that 
the two segments inside $\mycirc$ do not cross. Straight line segments of the drawing of the 
DFS-cycle will be called \emph{black segments}. Black segments inside $\mycirc$ will be called 
\emph{short black segments}, all the other black segments will be \emph{long black segments}.

\begin{figure}[hbt]
\begin{center}
	\epsfbox{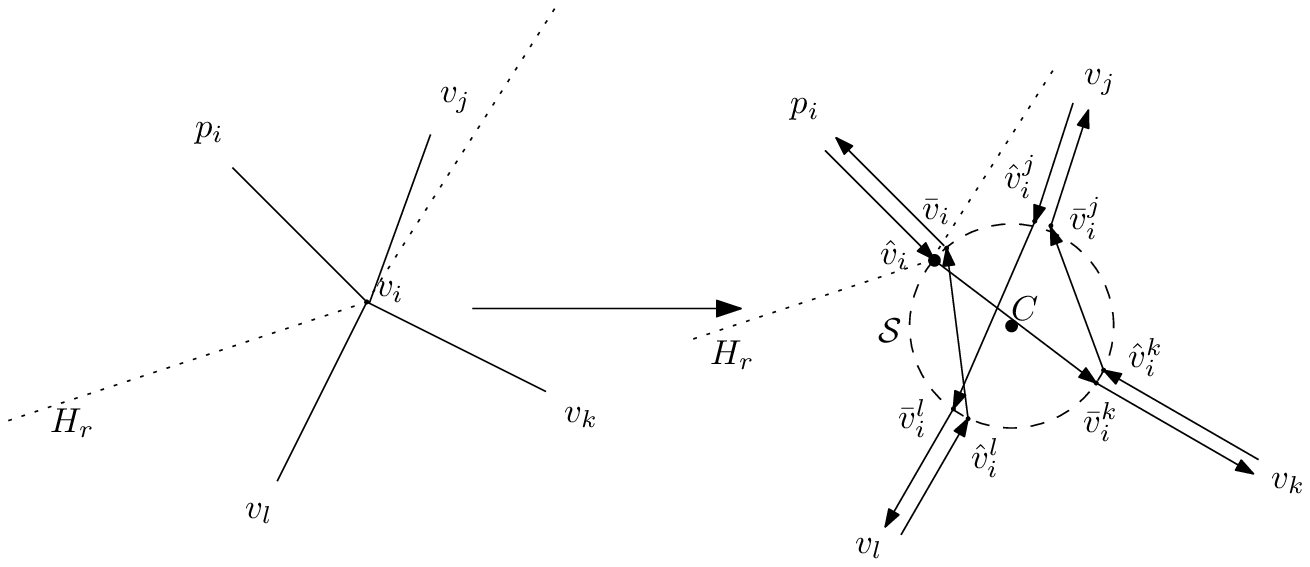}
	\caption{expanding the vertex $v_i$ in a plane drawing of $T'$}
	\label{fig_tree_expand}
\end{center}
\end{figure}

An edge $(u_i, u_{i+1})$ of $\hb$ is then drawn as the part of 
the drawing of the DFS-cycle between $\hat{u_i}$ and $\hat{u}_{i+1}$. 
Since $T'$ is the smallest subtree of $T$ containing all the fixed vertices, all the leaves
of $T'$ are fixed vertices. After visiting a leaf, a DFS visits another leaf after no more 
than $\diam(T)$ steps and thus there are at most $\diam(T)$ edges between two consecutive 
first visits of fixed vertices in the DFS cycle. Therefore the drawing of the whole 
cycle $\hb$ is composed of at most $\min\{4n,~2t\diam(T) \}$ black segments.

Because $\hr$ is drawn as a convex polygon, each straight line segment of $\hb$ crosses it 
in at most two points (after perturbing ends of the segments that have a common subsegment 
with $\hr$). There are thus at most $ \min\{8n,~4t\diam(T) \}$ crossings of pairs of edges 
with different colors.

Since $T'$ is drawn with no crossings, there are no crossings between any long black segment and 
another black segment. Short black segments cross only short black segments created by expanding
the same vertex of $T'$; but only if the degree of the vertex is at least three. 
Because $T'$ has at most $t$ leaves, there are at most $3t$ short black 
segments that cross some other short black segment. Each vertex of $T'$ has degree at most 
$\Delta$ and thus the number of crossings of pairs of black edges is at most $1.5 t \Delta$.

Let
\begin{eqnarray*}
K &:=& \l\lfloor 1.5 t \Delta + \min\{8n,2t \cdot \diam(T)\} \r\rfloor, \\
t &:=& \l\lceil 300 \sqrt{n} \log n
(\sqrt{\Delta} + \min\{\sqrt[6]{n/\log^2 n},~ \sqrt{\diam} \} ) \r\rceil,\\
s &:=& \l\lceil  35^2\frac{K+t}{t} \log^2 n \r\rceil .
\end{eqnarray*}

We can verify that $s<t$ as required.
We also have
\[
D = \l\lfloor 35\sqrt{\frac{t(K+t)}{s}}  \r\rfloor \leq \frac{t}{\log n}.
\]

If $T$ can be untangled while keeping $t$ vertices fixed, then for some $t$-tuple of its $n$ 
vertices, the above-defined graph $H$ has crossing number at most $K$. For every $t$-tuple, 
$\hb$ is exactly as required by Lemma~\ref{lemma_pt} 
and $\hr$ goes through the vertices of $\hb$ in a uniformly distributed random order, 
because at the beginning, we placed the vertices of $T$ in a uniformly distributed random order.

Therefore we can apply Lemma~\ref{lemma_pt} and
\begin{eqnarray*}
\prob[\fix(T,\delta) \geq t]   &\leq&  \binom{n}{t}\prob[cr(H) \leq K] \\
&\leq& \binom{n}{t} \binom{t}{D}^2 \l(\frac{2t}{s}\r)^D \frac{s^{t-D}}{(t-D)!} \\
&\leq& \l(\frac{en}{t}\r)^t \l(\frac{et}{D}\r)^{2D} \l(\frac{2t}{s}\r)^D \l(\frac{es}{t(1-\frac{1}{\log n})}\r)^{t-D} \\
&\leq& \l(e^2 \frac{ns}{t^2}\r)^t \l(\frac{2et^4}{D^2s^2}\r)^D  \l(1-\frac{1}{\log n}\r)^{D-t} \\
&\leq& \l(e^2 \frac{ns}{t^2}\r)^t \l(\frac{2e^3 t^4}{D^2s^2}\r)^{\frac{t}{\log n}}  \\
&\leq& \l(4e^2 \frac{ns}{t^2}\r)^t \\
&\leq& \l(8e^2 36^2 \frac{ n \log^2{n} (\min\{4n,t \cdot \diam(T)\} + t\Delta) } {t^3}\r)^t. \\
\end{eqnarray*}

We distinguish two cases:
\begin{enumerate}[(a)]

\item If $\sqrt[6]{n/\log^2 n} \geq \sqrt{\diam(T)}$, then
$t = \l\lceil 300 \sqrt{n} \log n (\sqrt{\Delta} + \sqrt{\diam(T)}) \r\rceil $ and

\begin{eqnarray*}
\prob[\fix(T,\delta) \geq t]   
&\leq& \l(8e^2 36^2 \frac{ n \log^2 n (\Delta + \diam(T)) } {t^2}\r)^t \\
&\leq& \l(\frac{8e^2 36^2}{300^2} \frac{ \Delta + \diam(T) } {(\sqrt{\Delta} + \sqrt{\diam(T)})^2}\r)^t \\
&<& 0.9^t.
\end{eqnarray*}

\item If $\sqrt[6]{n/\log^2 n} \leq \sqrt{\diam(T)}$, then
$t = \l\lceil 300 \sqrt{n} \log n (\sqrt{\Delta} + \sqrt[6]{n/\log^2 n} \r\rceil $ and
\begin{eqnarray*}
\prob[\fix(T,\delta) \geq t]   
&\leq& \l(8e^2 36^2 \frac{ n \log^2{n} (t\Delta + 4n) } {t^3}\r)^t \\
&\leq& \l(\frac{8e^2 36^2}{300^2} 
\frac{ t\Delta + 4n } {t \l( \sqrt{\Delta} + n^{\frac16} (\log n)^{-\frac13} \r)^2}\r)^t \\
&\leq& \l(\frac{8e^2 36^2}{300^2} 
\frac{ t\Delta + 4n } {t\Delta + t n^{\frac13} (\log n)^{-\frac23}}\r)^t \\
&\leq& \l(0.9 \frac{ t\Delta + 4n } {t\Delta + 300n^{\frac23}(\log n)^{\frac23} n^{\frac13} (\log n)^{-\frac23}}\r)^t \\
&<& 0.9^t.
\end{eqnarray*}

\end{enumerate}

Using the probabilistic method we conclude that for some drawing $\delta$ of $T$ it 
holds that $\fix(T,\delta) < t$. 
\end{proof}

\begin{corollary}
{\ }\par
\begin{enumerate}
\item
\label{item_all}
For every planar graph $G$ with maximum degree $\Delta$ and diameter $\diam$,
\[
\fix(G) \leq 300 \sqrt{n} \log n \l(\sqrt{\Delta} + \min\l\{\sqrt[6]{n/\log^2 n},~ \sqrt{2\diam} \r\}\r) .
\]

\item
\label{item_3conn}
There is a constant $c$ such that for every 3-vertex-connected planar graph $G$, 
\[
\fix(G) \leq c(n\log n)^{\frac23} .
\]

\item
\label{item_largedeg}
There is a constant $c$ such that for every constant $\epsi$, every planar graph $G$ on $n$ 
vertices with $\fix(G) \geq \epsi n$ has a vertex with degree at least
\[
\Delta \geq  c\frac{n \epsi^2}{\log^2 n} .
\]

\item
\label{item_smallfix}
There is a constant $c$ such that for every constant $b$, every planar graph $G$ on $n$ 
vertices with both maximum degree and diameter at most $b \log n$ satisfies
\[
\fix(G) \leq c \sqrt{bn\log^3 n}.
\]
\end{enumerate}
\end{corollary}

\begin{proof}
All claims are based on the simple observation, that adding edges to a graph
$H$ never increases $\fix(H)$ and thus if $T$ is a spanning tree of $G$, then $\fix(G) \leq \fix(T)$.
Part~\ref{item_largedeg} is then straightforward and part~\ref{item_3conn} follows from a theorem
of Barnette~\cite{Barnette1966} which says that every planar 3-vertex-connected graph has a spanning tree 
with maximum degree three. 

To prove parts~\ref{item_all} and \ref{item_smallfix}, we now show that any graph $G$ has a spanning 
tree $T$ with diameter at most $2\diam(G)$. Fix any vertex $v$ of $G$ and run a breadth-first search 
from it. 
All vertices lie at distance at most $\diam(G)$ from $v$ and thus the diameter of the breadth-first search
tree is at most $2 \diam(G)$.
\end{proof}

{\bf Acknowledgments.}  I am grateful to Alexander Wolff for a careful
reading of an earlier version of the paper and many useful suggestions.

\bibliographystyle{plain}

\bibliography{untangling}

\end{document}